\documentclass[11pt]{article}

\usepackage{fullpage}
\usepackage[ruled, linesnumbered, vlined, commentsnumbered]{algorithm2e}
\usepackage{graphicx,subfig} 
\usepackage{amsfonts,amssymb,amsmath,amsthm,amsopn,mathtools}	
\usepackage{booktabs,diagbox,colortbl,multirow,tabularx,threeparttable,hhline}
\usepackage[listings,skins,breakable]{tcolorbox}

\usepackage{enumerate}
\usepackage{authblk}
\usepackage{footnote}
\usepackage{hyperref}
\usepackage{prettyref}
\usepackage{cite}
\usepackage{setspace}
\usepackage{color}
\usepackage{xcolor}  
\usepackage{scrpage2}
\usepackage{geometry}

\usepackage{tikz}
\usepackage{pgfplots}
\usetikzlibrary{positioning,shapes,shadows,arrows,calc}
\tikzstyle{component}=[rectangle, draw=black, rounded corners, fill=blue!40, drop shadow, text centered, anchor=north, text=white, minimum height=1cm]
\tikzstyle{arrow}=[->, thick]

\pgfplotsset{compat=1.12}
\usetikzlibrary{intersections}
\usetikzlibrary{pgfplots.statistics}
\usepgfplotslibrary{fillbetween}

\definecolor{myblue}{RGB}{34,31,217}
\definecolor{mycyan}{gray}{.7}
\definecolor{Gray}{gray}{0.9}

\DeclareMathOperator*{\argmax}{argmax}

\newcommand{\pref}{\prettyref}

\newrefformat{fig}{Fig.~\ref{#1}}
\newrefformat{tab}{Table~\ref{#1}}
\newrefformat{sec}{Section~\ref{#1}}
\newrefformat{alg}{Algorithm~\ref{#1}}
\newrefformat{property}{Property~\ref{#1}}
\newrefformat{theorem}{Theorem~\ref{#1}}
\newrefformat{definition}{Definition~\ref{#1}}
\newrefformat{corollary}{Corollary~\ref{#1}}
\newrefformat{lemma}{Lemma~\ref{#1}}
\newrefformat{conj}{Conjecture~\ref{#1}}
\newrefformat{def}{Definition~\ref{#1}}
\newrefformat{eq}{equation~(\ref{#1})}
\newrefformat{app}{Appendix~\ref{#1}}

\usepackage{lscape}

\begin{document}

\title{\vspace{-1ex}\LARGE\textbf{Art-Attack: Black-Box Adversarial Attack via Evolutionary Art}~\footnote{This manuscript is submitted for potential publication. Reviewers can use this version in peer review.}}

\author[1]{\normalsize Phoenix Williams}
\author[2]{\normalsize Ke Li}
\affil[1,2]{\normalsize Department of Computer Science, University of Exeter, EX4 4QF, Exeter, UK}
\affil[1]{\normalsize Email: \texttt{pw384@exeter.ac.uk}}
\affil[2]{\normalsize Email: \texttt{k.li@exeter.ac.uk}}

\date{}
\maketitle

\vspace{-3ex}
{\normalsize\textbf{Abstract: } Deep neural networks (DNNs) have achieved state-of-the-art performance in many tasks but have shown extreme vulnerabilities to attacks generated by adversarial examples. Many works go with a white-box attack that assumes total access to the targeted model including its architecture and gradients. A more realistic assumption is the black-box scenario where an attacker only has access to the targeted model by querying some input and observing its predicted class probabilities. Different from most prevalent black-box attacks that make use of substitute models or gradient estimation, this paper proposes a gradient-free attack by using a concept of evolutionary art to generate adversarial examples that iteratively evolves a set of overlapping transparent shapes. To evaluate the effectiveness of our proposed method, we attack three state-of-the-art image classification models trained on the CIFAR-10 dataset in a targeted manner. We conduct a parameter study outlining the impact the number and type of shapes have on the proposed attack's performance. In comparison to state-of-the-art black-box attacks, our attack is more effective at generating adversarial examples and achieves a higher attack success rate on all three baseline models.}

{\normalsize\textbf{Keywords: } }Adversarial Examples, Deep Learning, Computer Vision, Evolutionary-art


\section{Introduction}
\label{sec:introduction}

Deep neural networks (DNNs) have achieved state-of-the-art performance in various tasks and have been successfully applied in real-world problems such as autonomous driving~\cite{BojarskiTDFFGJM16}, identity verification~\cite{LiuLLSG18} and financial services~\cite{BrigoHPB21}. Despite their successes, DNNs have exhibited extreme vulnerabilities to adversarial examples~\cite{GoodfellowSS14,SzegedyZSBEGF13,KurakinGB16,PapernotMG16}. By adding a small perturbation to a correctly classified image, adversarial examples have been shown to cause trained DNNs to misclassify. It has also been shown that specific adversarial examples can cause a DNN to misclassify to a particular label~\cite{AlzantotSCS2019, RuCBG20}. Such adversarial examples have been proven to exist in the physical world~\cite{KurakinGB16} as well as in different domains such as malware detection~\cite{Hu018} and code naturalness~\cite{Yefet0Y20}. Due to the notorious lack of robustness of DNNs to adversarial examples, particular concerns have been raised for security-critical applications~\cite{AlzantotSCS2019}.
It has been pointed out that one of the keys to addressing this issue is finding strong adversarial examples~\cite{BaiL0WW21} which have sparked a growing interest in finding such examples using attack algorithms. Many existing works assume a full access to the targeted model, e.g., the gradient information a.k.a. \textit{white-box} attacks~\cite{SzegedyZSBEGF13, GoodfellowSS14, KurakinGB16, MadryMSTV17, Carlini017, ChenSZYH18, SharmaC17}. However, consideration must also be made for the \textit{black-box} setting where the attacker has no prior information of the targeted model and can only interact by querying an input and observing the model's predicted class probabilities. A popular \textit{black-box} adversarial attack approach is to train a \textit{substitute} model and carry out a \textit{white-box} adversarial attack hoping the generated adversarial examples also deceive the target model~\cite{PapernotMG16, GuoYZ19, ChengDPSZ19}. A core issue with this approach is the assumption that the original training data is available to train the \textit{substitute} model. Additionally, the inherent computational cost of training the model on a possibly large training data-set means it is unlikely to be used in a real-world scenario. Alternatively, another class of \textit{black-box} attacks estimate the gradient of the objective function \ref{eq:targettedAttack} or \ref{eq:untargettedAttack} using methods of finite-differences~\cite{ChenZSYH17,TuTC0ZYHC19,IlyasEAL18, BhagojiHLS18,UesatoOKO18}. However, due to the high dimension of the image search space, such works require a large number of model queries to estimate the gradient. Another direction of \textit{black-box} attacks remove the need for any gradient estimation by applying random search~\cite{CroceASFH20} or evolutionary algorithms~\cite{AlzantotSCS2019, MeunierAT19}. The large majority of works addressing \textit{black-box} adversarial attacks do not consider the perturbation to have any structure and aim to optimize every pixel, applying bilinear interpolation or embedding mechanisms to deal with a high dimensional search space. 
\textbf{Contributions.} Motivated by the previous work outlined above we proposed a novel approach that generates adversarial examples under a \textit{black-box} setting. Our proposed attack does not rely on any gradient information and only assumes access to the targeted model by querying images and observing predicted class probabilities. To perform a \textit{gradient-free} attack, we adopt an evolutionary-art algorithm that iteratively evolves a single solution of shapes until the attack is successful or the maximum number of target model queries is reached. We analyse the impact the number and type of used shapes have on the proposed algorithm in terms of its effectiveness and efficiency in attacking image classification models.
To evaluate our proposed attack in comparison with peer-algorithms in the literature we conduct thorough \textit{targeted} attacks on three image classification models~\cite{SimonyanZ14a, LinCY13, SpringenbergDBR14} trained on the CIFAR-10 data set.
We make use of the models proposed in~\cite{SimonyanZ14a, LinCY13, SpringenbergDBR14} as they are widely used in the literature. Our algorithm achieves higher attack success rates than current state-of-the-art black-box attacks in the literature with perturbation limitations.
In summary, we make the following contributions:
\begin{itemize}
    \item We propose a novel gradient-free approach that generates adversarial examples by iteratively evolving a single solution of transparent RGB valued shapes placed on a targeted image.
    \item Without requiring the use of dimension reduction, we show that our algorithm can generate adversarial examples even when restricting the size of the perturbation.
    \item We highlight the effectiveness of the proposed attack by outperforming state-of-the-art \textit{black-box} attacks on three state-of-the-art image classification models. 
\end{itemize}
The rest of this paper is organized as follows. \pref{sec:preliminary} provides a summary of related works and the attack setting we address in this work. \pref{sec:threatmodel} describes the objective function used within our attack. \pref{sec:ProposedAlg} outlines the proposed attack and gives details of its implementation and adversarial image construction mechanism. \pref{sec:experiments} describes our evaluation of the proposed as well as the conducted parameter study. We conclude this paper in~\pref{sec:conclusion} with a discussion of the proposed attack and directions for our future work. 


\section{Preliminary}
\label{sec:preliminary}
Let $f: \mathcal{X} \subseteq \mathbb{R}^{d} \rightarrow \mathbb{R}^{K}$ be a classifier which takes an input $x \in \mathcal{X}$ and assigns it a class $y = \argmax{r=1,\cdots,K}{f_{r}(x)}$ where $f_{r}$ is the probability of input $x$ being of class $r$. A \textit{targetted} attack on such a model involves finding a perturbation $\delta \in \mathbb{R}^{d}$ such that

\begin{equation}
\begin{aligned}
\argmax{r=1,\cdots,K}{f_{r}(x + \delta)}=c, \hspace{2mm} \delta \in \mathcal{T} \text{ and } x + \delta \in \mathcal{X}
\end{aligned}
\label{eq:targettedAttack}
\end{equation}
where $\mathcal{T}$ are the constraints on the perturbation $\delta$, $\mathcal{S}$ is the input domain and $c$ is the target label which is not its correctly predicted class. \textit{Untargeted} attacks however aim to find a perturbation $\delta$ such that

\begin{equation}
\begin{aligned}
\argmax{r=1,\cdots,K}{f_{r}(x + \delta)} \neq y, \hspace{2mm} \delta \in \mathcal{T} \text{ and } x + \delta \in \mathcal{X}
\end{aligned}
\label{eq:untargettedAttack}
\end{equation}
where $y$ is the correctly predicted class of the input $x$. The discovery of a $\delta$ that satisfies \ref{eq:targettedAttack} can be described by the optimization of
\begin{equation}
    \begin{aligned}
    \max{\delta \in \mathbb{R}^{d}}L(f(x+\delta), t_{target}), \hspace{2mm} \delta \in \mathcal{T} \text{ and } x + \delta \in \mathcal{X}
    \end{aligned}
    \label{eq:TargettedAttackOpt}
\end{equation}
where the maximization of $L$ leads to the desired classification.
Many algorithms have been proposed to solve \ref{eq:TargettedAttackOpt}, in the white-box setting ~\cite{SzegedyZSBEGF13, GoodfellowSS14, KurakinGB16, MadryMSTV17, Carlini017, ChenSZYH18, SharmaC17} it is assumed that the attacker has full access to the targeted classifier. Using back-propagation for gradient computation the attacker structures the attack as a gradient-based optimization problem. In this work, we are concerned with attacking a classifier within the score-based black-box setting where only the probabilities of each class returned by the classifier are accessible. Work addressing this problem in the literature aim to find adversarial examples which are constrained by their $l_{0}, l_{2}$ or $l_{\infty}$ norms. Generating adversarial examples which are constrained by $||x_{adv}-x||_{0} \leq \epsilon$ involves corrupting $n \leq \epsilon$ individual pixels of the original image. Notable papers ~\cite{SuVS19, CroceASFH20} have adapted previously proposed optimization algorithm to the $l_{0}$ adversarial attack setting. In particular, Su et. al ~\cite{SuVS19} adapt the popular differential evolution ~\cite{StornP97} algorithm by allowing evolved solutions to represent the $x,y$ coordinates of each pixel and their corresponding RGB value. Similarly, Croce et. al ~\cite{CroceASFH20} propose a random search algorithm that iteratively samples a distribution of pixel index's and perturbations to generate new solutions. 
$l_{2}$ and $l_{\infty}$ constrained score-based adversarial attacks aim to find adversarial examples that satisfy $||x_{adv} - x||_{2} \leq \epsilon$ and $||x_{adv} - x||_{\infty} \leq \epsilon$ respectively. Works addressing these settings in the literature have made use of finite-differences~\cite{ChenZSYH17, TuTC0ZYHC19, IlyasEAL18, BhagojiHLS18, UesatoOKO18} for gradient estimation and solve \pref{eq:TargettedAttackOpt} using gradient methods. In particular, the ZOO~\cite{ChenZSYH17} algorithm proposed by Chen et. al makes use of the Adam~\cite{KingmaB14} optimization algorithm to generate adversarial examples. A large issue with this approach is the number of model queries required to compute the gradient of a solution within the high-dimensional image space. To alleviate the high dimensional nature, works have made use of dimension reduction techniques such as bi-linear interpolation~\cite{AlzantotSCS2019, ChenZSYH17, TuTC0ZYHC19, IlyasEAL18} and the use of auto-encoders~\cite{TuTC0ZYHC19}. To further the query efficiency works in the literature~\cite{IlyasEAL18, TuTC0ZYHC19} have estimated the gradients by sampling zero-mean and unit-variance normal distributions. Other works include the use of evolutionary algorithms and random search to attack classifiers~\cite{AlzantotSCS2019,MeunierAT19,GuoGYWW19}, removing the need for gradient estimation and have shown an ability to outperform gradient estimate algorithms in the literature. In particular, the proposed \textit{GenAttack} of~\cite{AlzantotSCS2019} evolves a population of adversarial images using genetic operators and applies adaptive parameter scaling and bi-linear interpolation to handle the high-dimension of the search space. Attacks based on Bayesian optimization have also been proposed~\cite{ShuklaSWK19, RuCBG20} and have shown good performance in the low-query availability setting.
Another approach to black-box adversarial attack is the use of a substitute model~\cite{PapernotMG16,GuoYZ19,ChengDPSZ19} where the attacker has access to a model similar to the target model. By attacking the substitute model a series of adversarial examples are generated which are used to attack the original target model. In this work we do not assume access to such a similar model and therefore do not compare with works addressing this scenario. Other addressed scenarios include the setting where accessible information is limited to the probabilities or labels of the top $k$ classes ~\cite{IlyasEAL18}. This setting is a generalization of the \textit{decision-based} attack ~\cite{BrendelRB18, ChengLCZYH19, GuoFW19, BrunnerDTK19} where only $k=1$ labels are returned by the model.

\section{Threat Model}
\label{sec:threatmodel}
Given a target model $f:\mathcal{X} \subseteq \mathbb{R}^{d} \rightarrow \mathbb{R}^{K}$ that takes an input $x \in [0,1]^{d}$, we assume the attacker only has access to its predicted probabilities by querying $f$ with an input $x_{adv}$. To this end, we make use of the optimization problem employed by the GenAttack algorithm ~\cite{AlzantotSCS2019} that aims to maximize the probability of the target class $f_{c}(x_{adv})$ whilst jointly minimizing the probability of the other classes. Given an adversarial image $x_{adv}=x+\delta$ where $x$ is the attacked image, the optimization problem is described by
\begin{equation}
    \begin{aligned}
        \max{\delta \in \mathbb{R}^{d}}\log({f_{c}(x + \delta)}) -\log({\Sigma_{i \neq c} f_{i}(x + \delta)}), \hspace{1mm} 
        ||x_{adv} - x||_{\infty} \leq \epsilon
    \end{aligned}
    \label{eq:threatmodel}
\end{equation}
where $c$ is the target class and $\delta$ is constrained by $ \delta \in [-\epsilon, \epsilon]$. The log of the probabilities are used to overcome any numerical instabilities that may arise.

\section{Proposed Attack Method}
\label{sec:ProposedAlg}

\begin{figure}[!t]
    \centering
    \includegraphics[width=\linewidth]{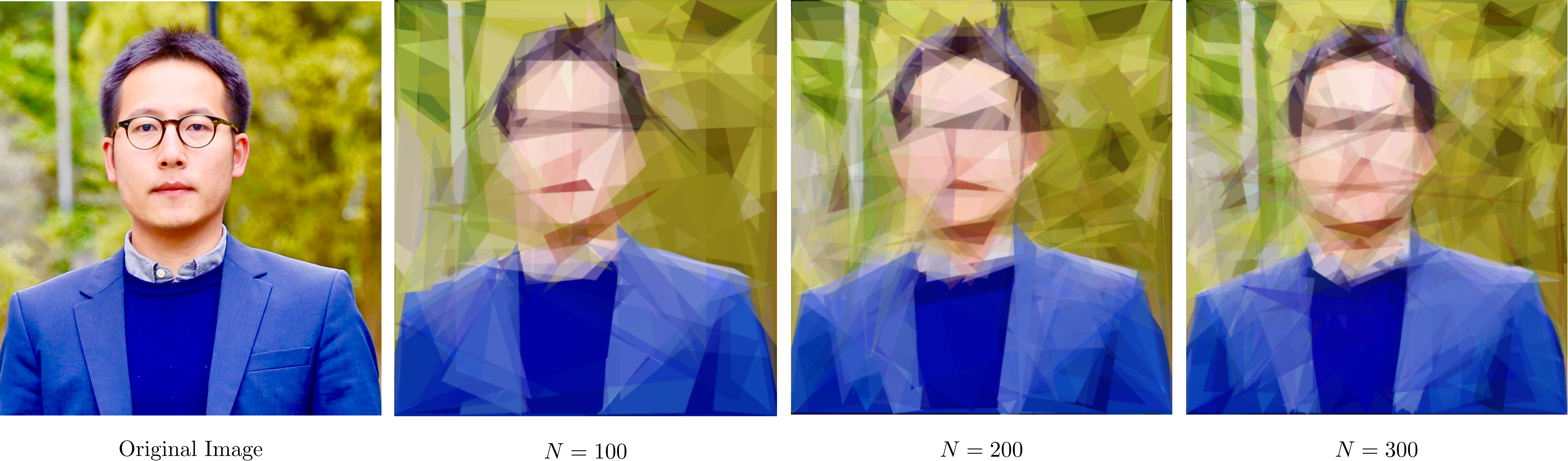}
    \caption{Evolutionary Art algorithm applied to image reconstruction with $N=100$, $N=200$ and $N=300$ polygons.}
    \label{fig:evokl}
\end{figure}
Evolutionary art algorithms apply evolutionary optimization algorithms to construct pieces of music ~\cite{Biles94, MoroniMZG00, Biles94} or art ~\cite{World96, LambertLL13, poli1997evolution}. In this work we employ an evolutionary art algorithm that evolves a single solution of $N$ shapes by iteratively applying a mutation operator onto the solution, figure \ref{fig:evokl} shows its results when applied to image reconstruction using 3-sided polygons. In each generation, a child solution is created by applying the mutation operator to the parent solution. Only the solution with the better objective function value is kept for the next generation. We represent a shape as a single-dimensional array of length $a$ which holds its properties such as position, colour and transparency. The length $a$ is dependent on the shape used. A single solution $y$ is represented as a collection of $N$ shapes which forms a matrix $y \in \mathbb{R}^{N \times a}$. 

\textbf{Circle Objective Function}: For a solution $y \in \mathbb{R}^{N \times a}$ where each $y_{i} \in \mathbb{R}^{7}$ represents a circle and $y_{ij}$ is bounded within $[0, 1]$ we generate an adversarial image $x_{adv}$ that satisfies the constraints $||x_{adv} - x||_{\infty} \leq \epsilon$. For this we iteratively place each circle $y_{i}$ onto the attacked image $x$. The values $y_{i0}$ and $y_{i1}$ describe the center of the $ith$ circle by multiplying with the image's height and width respectively and rounding to the closest integer, we represent this multiplication in algorithm \ref{alg:CircleObjectiveFunction} by $\hat{\times}$. Using this $\hat{\times}$ operator the radius of the $ith$ circle is described by $y_{i2}$. The color and transparency of each circle is expressed by the values $y_{i3}, y_{i4}, y_{i5}$ and $y_{i6}$ respectively. Once all circles have been layered onto the attacked image generating the adversarial image $x_{adv}$, we project $x_{adv}$ onto the space $\{\mathcal{X} \in \mathbb{R}^{d} : ||x_{adv} - x||_{\infty}|| \leq \epsilon\} \bigcap [0, 1]^{d}$ using the clipping mechanism and return the projected adversarial image $x_{adv}$. We describe the full process in algorithm \ref{alg:CircleObjectiveFunction}.

\begin{algorithm}[t]
    \KwIn{Solution $y$, Attacked image $x \in \mathbb{R}^{d} \bigcap [0, 1]^{d}$, $l_{\infty}$ constraint $\epsilon$, radius divider $\beta$}
    $h \leftarrow x.height, w \leftarrow x.width$ \\
    $r \leftarrow \frac{h+w}{\beta}$ \\
    $x_{adv} \leftarrow x$ \\
    \For{$y_{i}$ in $y$}
    {
    $center \leftarrow (y_{i1} \hat{\times} h, y_{i0} \hat{\times} w$) \\
    $radius \leftarrow y_{i2} \hat{\times} r$ \\
    $color \leftarrow (y_{i3}, y_{i4}, y_{i5})$ \\
    $\alpha \leftarrow y_{i6}$ \\
    $s \leftarrow Circle(center, radius, color, \alpha)$ \\
    $x_{adv} \leftarrow x_{adv} + s$ \\
    }
    Project $x_{adv}$ onto $\{\mathcal{X} \in \mathbb{R}^{d} : ||x_{adv} - x||_{\infty}|| \leq \epsilon\} \bigcap [0, 1]^{d}$ \\
    \Return $x_{adv}$
    \caption{Circle Adversarial Image Generation}
    \label{alg:CircleObjectiveFunction}
\end{algorithm}
\textbf{Triangle Objective Function}: Following a highly similar process to the circle objective function we use a solution $y \in \mathbb{R}^{N \times a}$ where each $y_{i} \in \mathbb{R}^{10}$ represents a triangle and $y_{ij}$ is bounded within $[0, 1]$ to generate an adversarial image. For each triangle $y_{i}$ we describe the positions of its vertices as $[[y_{i0} \hat{\times} h, y_{i1} \hat{\times} w]$,\\ $[y_{i2} \hat{\times} h, y_{i3} \hat{\times} w]$, $[y_{i4} \hat{\times} h, y_{i5} \hat{\times} w]]$ where $h$ and $w$ is the height and width of the attacked image $x$. The color and transparency of each triangle is expressed by the values $y_{i6}, y_{i7}, y_{i8}$ and $y_{i9}$ respectively. Before returning the adversarial image $x_{adv}$ we project it onto the space $\{\mathcal{X} \in \mathbb{R}^{d} : ||x_{adv} - x||_{\infty}|| \leq \epsilon\} \bigcap [0, 1]^{d}$. Algorithm \ref{alg:TriangleObjectiveFunction} described the full procedure.

\begin{algorithm}[t]
    \KwIn{Solution $y$, Attacked image $x \in \mathbb{R}^{d} \bigcap [0, 1]^{d}$, $l_{\infty}$ constraint $\epsilon$}
    $h \leftarrow x.height, w \leftarrow x.width$ \\
    $x_{adv} \leftarrow x$ \\
    \For{$y_{i}$ in $y$}
    {
    $vertices = 
    \begin{aligned}
        [[y_{i0} \hat{\times} h, y_{i1}  \hat{\times} w], \\
        [y_{i2} \hat{\times} h, y_{i3}  \hat{\times} w],  \\
        [y_{i4} \hat{\times} h, y_{i5} \hat{\times} w]]
    \end{aligned}$ \\ \vspace{1mm}
    $color \leftarrow (y_{i6}, y_{i7}, y_{i8})$ \\
    $\alpha \leftarrow y_{i9}$ \\
    $t \leftarrow Triangle(vertices, color, \alpha)$ \\
    $x_{adv} \leftarrow x_{adv} + t$ \\
    }
    Project $x_{adv}$ onto $\{\mathcal{X} \in \mathbb{R}^{d} : ||x_{adv} - x||_{\infty}|| \leq \epsilon\} \bigcap [0, 1]^{d}$ \\
    \Return $x_{adv}$
    \caption{Triangle Adversarial Image Generation}
    \label{alg:TriangleObjectiveFunction}
\end{algorithm}
\textbf{Rectangle Objective Function}: Given a single shape $y_{i}$ within a solution $y \in \mathbb{R}^{N \times a}$, we construct a square by first determining the positions of its top-left and bottom-right corners. This is done using the $\hat{\times}$ operator on the $y_{i1}, y_{i2}$ and $y_{i0}, y_{i3}$ entries with the targeted images width and height respectively. The color and transparency of the square are determined by $y_{i4}, y_{i5}, y_{i6}$ and $y_{i7}$ respectively. Once the square $y_{i}$ as been constructed it is placed onto the targeted image. Alike the previous objective functions, once the adversarial image has been constructed it is projected onto the space $\{\mathcal{X} \in \mathbb{R}^{d} : ||x_{adv} - x||_{\infty}|| \leq \epsilon\} \bigcap [0, 1]^{d}$ through a clipping mechanism. We describe this process in algorithm \ref{alg:SquareObjectiveFunction}.

\begin{algorithm}[t]
    \KwIn{Solution $y$, Attacked image $x \in \mathbb{R}^{d} \bigcap [0, 1]^{d}$, $l_{\infty}$ constraint $\epsilon$}
    $h \leftarrow x.height, w \leftarrow x.width$ \\
    $x_{adv} \leftarrow x$ \\
    \For{$y_{i}$ in $y$}
    {
    $corners = 
    \begin{aligned}
        [[y_{i0} \hat{\times} h, y_{i1} \hat{\times} w], \\
        [y_{i3} \hat{\times} h, y_{i2} \hat{\times} w]]
    \end{aligned}$ \\ \vspace{1mm}
    $color \leftarrow (y_{i4}, y_{i5}, y_{i6})$ \\
    $\alpha \leftarrow y_{i7}$ \\
    $s \leftarrow Square(corners, color, \alpha)$ \\
    $x_{adv} \leftarrow x_{adv} + t$ \\
    }
    Project $x_{adv}$ onto $\{\mathcal{X} \in \mathbb{R}^{d} : ||x_{adv} - x||_{\infty}|| \leq \epsilon\} \bigcap [0, 1]^{d}$ \\
    \Return $x_{adv}$
    \caption{Rectangle Adversarial Image Generation}
    \label{alg:SquareObjectiveFunction}
\end{algorithm}
\textbf{Mutation Operator}: Given a solution $y \in \mathbb{R}^{N \times a}$ of $N$ shapes, the mutation operator applies random changes to generate a child solution. The first stage of the mutation operator sets $y_{child}$ as a copy of $y$ and generates two random integers $c$ and $change$ within the bounds $[0, N-1]$ and $[0, a+1]$ respectively. With a probably of $1/a$ the mutation operator randomly rolls $y_{child}$ solution between the index's $c$ and $j$ which is assigned a random integer value within the bound $[0, N-1]$. For instance, given a solution $x = \{s_{1}, s_{2}, ...s_{N-1}, s_{N}\}$ an output of the roll function with $shifts=2$ is $x_{roll} = \{s_{N-1}, s_{N}, s_{1}, s_{2},.., s_{N-3}, s_{N-2}\}$ whereas setting $shifts=-2$ would produce $x_{roll}=\{s_{3}, s_{4},...,s_{N-1}, s_{N}, s_{2}, s_{1}\}$. The second stage of the operator generates a list of random indices of length $change$. These indices with the previously generated $c$ determines which values we amend. With a probability of $\mu$ the values are set to random samples from a uniform distribution within the range $(0, 1)$. Otherwise, the values are amended by adding values sampled from a uniform distribution within the range $(-0.5, 0.5)$. Finally, the mutation operator projects the $y_{child}$ solution onto the valid space $[0,1]$ using a clipping mechanism. We describe this process in algorithm \ref{alg:Mutation}.

\begin{algorithm}[t]
    \KwIn{Solution $y$, Number of shapes $N$, shape array length $a$, mutation rate $\mu$}
    $c \leftarrow RandInt(0, N-1)$\\
    $change \leftarrow RandInt(0, a+1)$ \\
   $y_{child} \leftarrow y$ \\
    \If{$change \geq a+1$}{
        $change \leftarrow change - 1$ \\
        $i \leftarrow c, j \leftarrow RandInt(0, N-1)$\\
        \If{i < j}{
            $i \leftarrow i, j \leftarrow j, s \leftarrow -1$\\
        }
        \Else{
            $i \leftarrow j, j \leftarrow i, s \leftarrow 1$\\
        }
        $y_{child_{i: j+1}} \leftarrow Roll(y_{child_{i: j+1}} , shifts=s)$\\
    }
    $idx \leftarrow RandChoice(list[0: a], size=change)$ \\
    \If{$\mathcal{U}(0, 1)$ < $\mu$}
    {
        $y_{child_{c: idx}} \leftarrow \mathcal{U}(0, 1)$
    }
   \Else
    {
       $y_{child_{c: idx}} \leftarrow y_{child_{c: idx}} + \mathcal{U}(-0.5, 0.5)$
    }
    Project $y_{child}$ onto $\{\mathcal{Y} \in \mathbb{R}^{N \times a}\} \bigcap [0, 1]^{N times a} $\\
    \Return $y_{child}$
    \caption{Mutation Operator}
    \label{alg:Mutation}
\end{algorithm}

\begin{algorithm}
\KwIn{Classifier objective function $L$, attacked image $x \in \mathbb{R}^{d}$, $l_{\infty}$ radius $\epsilon$, Target label $c$, maximum number of iterations $M$, number of shapes $N$, shape array length $a$}
\vspace{1mm}
$y \leftarrow \mathcal{U}(0, 1)^{a \times N}$ // \textit{Uniform Initialization} \\
$l \leftarrow L(y, c)$,  $i \leftarrow 1$ \\
\While{$i<M$}
{
    $y_{new} \leftarrow Mutation(y)$ // see Alg. \ref{alg:Mutation} \\
    $l_{new} \leftarrow L(y_{new}, c)$ // see Alg. \ref{alg:CircleObjectiveFunction}, \ref{alg:TriangleObjectiveFunction} and \ref{alg:SquareObjectiveFunction} for Obj. functions \\
    
    \If{$l_{new} > l$}
    {
        $y \leftarrow y_{new}$ \\
        $l \leftarrow l_{new}$ \\
    }
    $i \leftarrow i + 1$
    
    \If{$y$ \textit{is a successful adversarial image}}
    {
        \Return $y$ // Successful adversarial image
    }
    
}
\caption{The Evolutionary-Art Attack Algorithm}  
\end{algorithm}

\section{Experiments}
\label{sec:experiments}

\begin{figure}[!t]
    \centering
    \includegraphics[width=\linewidth]{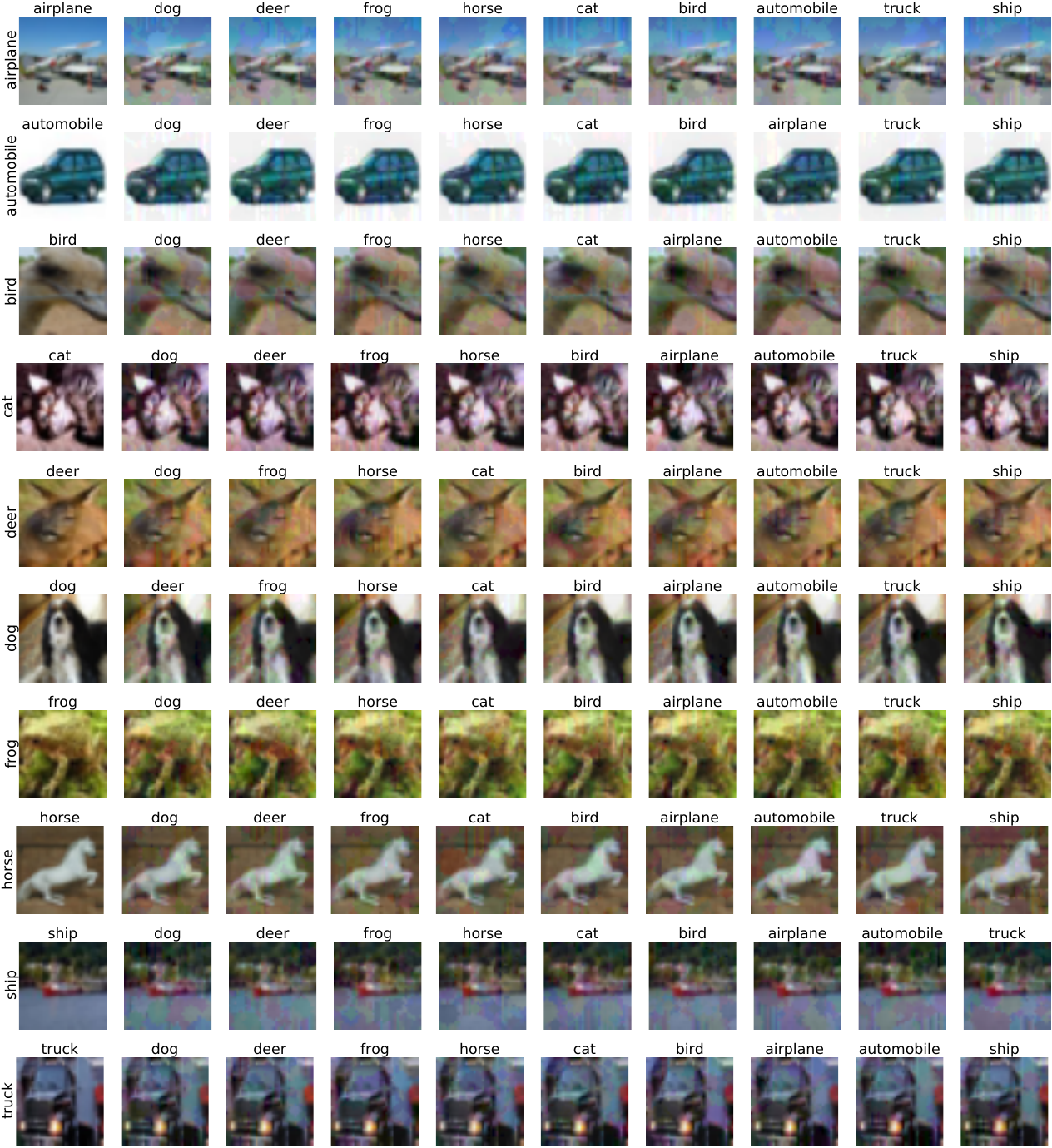}
    \caption{Adversarial Images generated by attacking a classification model ~\cite{SpringenbergDBR14} with 100 Circles ($\epsilon=0.05$). Row label is the true label and column label is the target label.}
    \label{fig:CNNcircle100}
\end{figure}

\begin{figure}[!t]
    \centering
    \includegraphics[width=\linewidth]{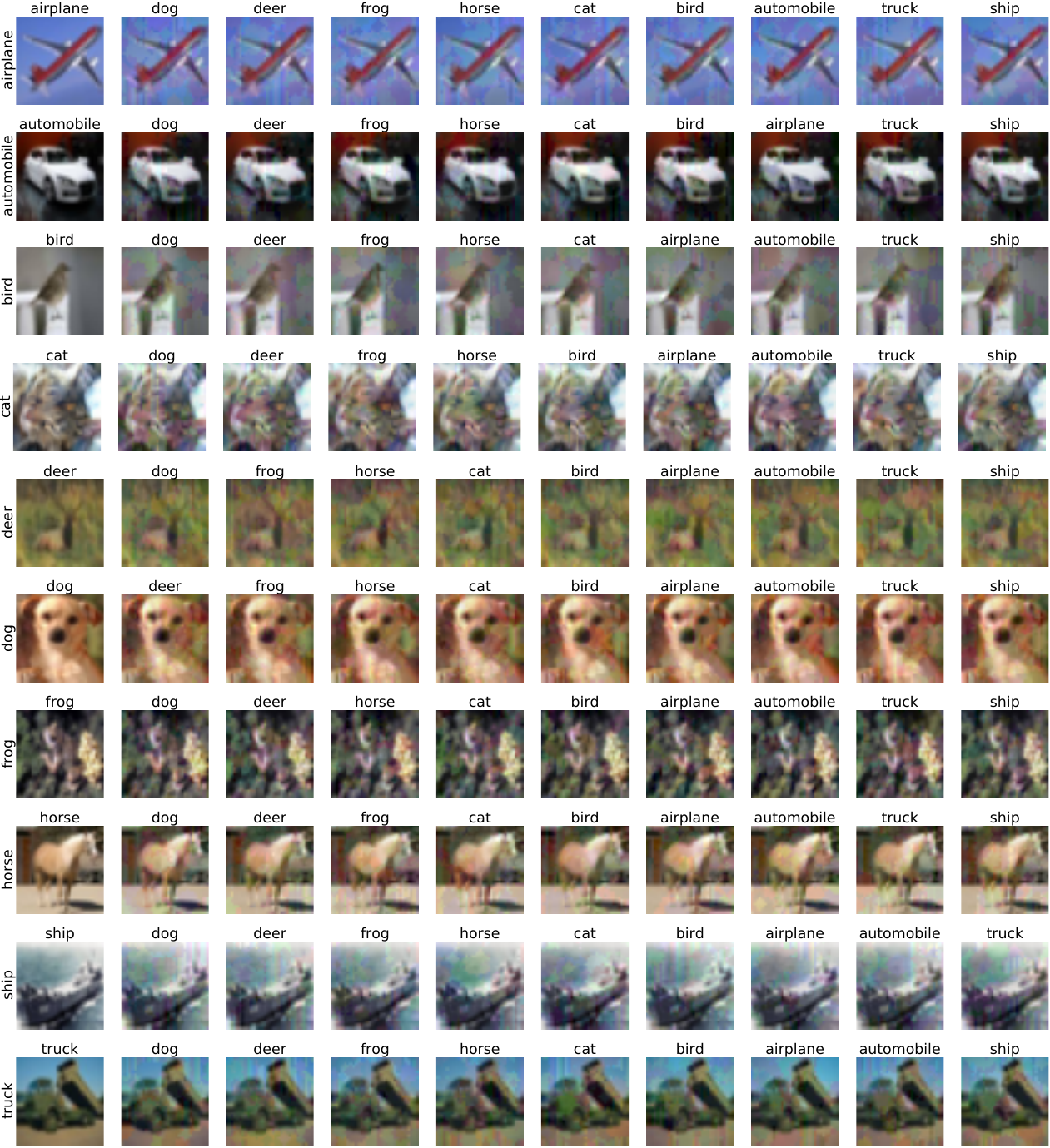}
    \caption{Adversarial Images generated by attacking a classification model model ~\cite{SimonyanZ14a} with 100 Circles ($\epsilon=0.05$). Row label is the true label and column label is the target label.}
    \label{fig:VGG16circle100}
\end{figure}

\begin{figure}[!t]
    \centering
    \includegraphics[width=\linewidth]{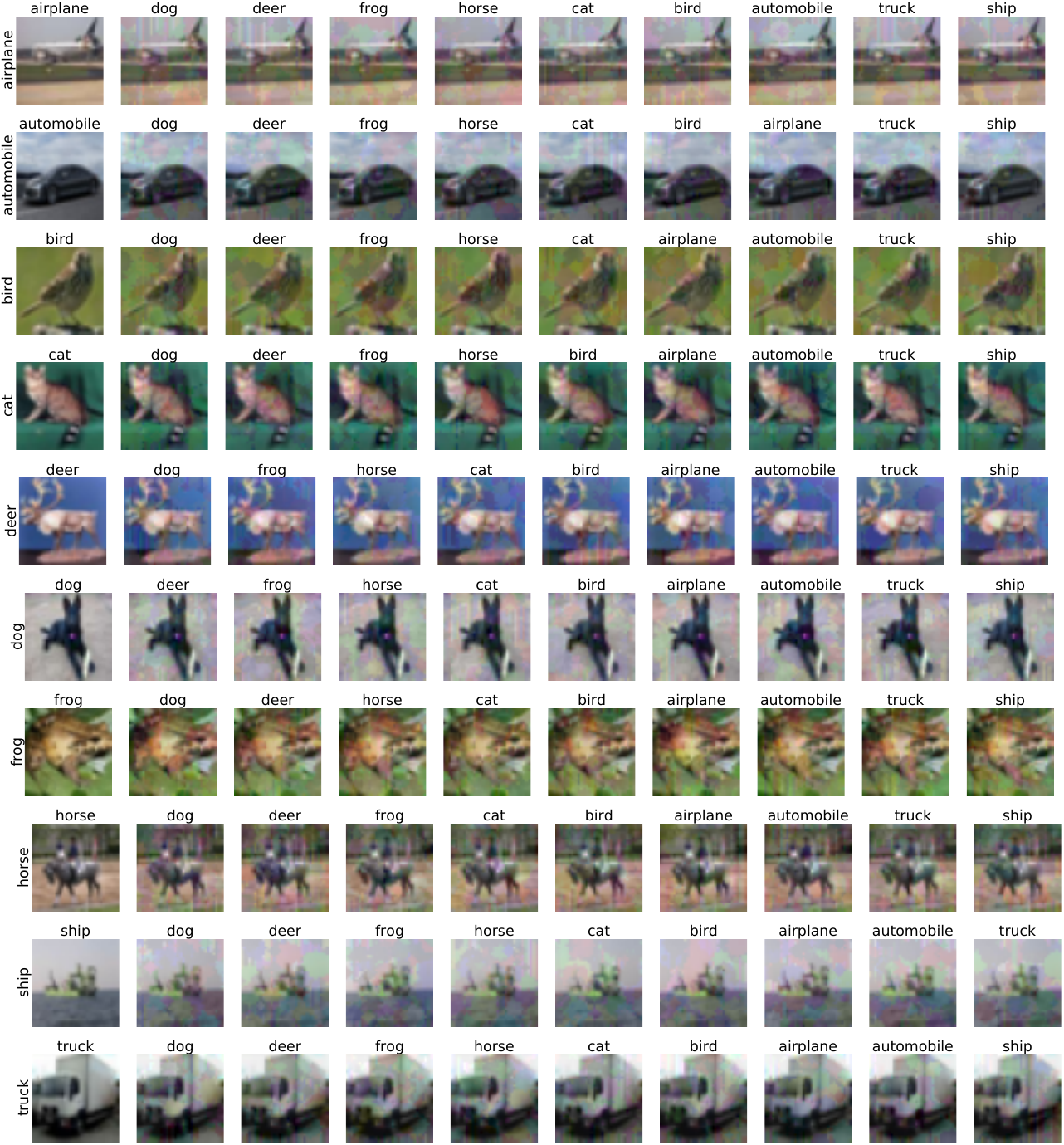}
    \caption{Adversarial Images generated by attacking a classification model ~\cite{LinCY13} with 100 Circles ($\epsilon=0.05$). Row label is the true label and column label is the target label.}
    \label{fig:NiNcircle100}
\end{figure}

In this section, we empirically demonstrate the effectiveness of the proposed attack by conducting \textit{targeted attacks} with $\epsilon=0.05$ on three classification models ~\cite{LinCY13, SimonyanZ14a, SpringenbergDBR14} trained on the CIFAR-10 data set, the models achieve test set accuracies of $85\%, 93\%$ and $87\%$ respectively. The reader can refer to ~\cite{SuVS19} for the full architectural details of the models. Evaluating the proposed attack we first conduct a parameter study analysing the impact of varying the type and number of shapes have on the performance of the proposed attack. Concluding this study we compare the effectiveness of the proposed algorithm with that of AutoZoom ~\cite{TuTC0ZYHC19}, GenAttack ~\cite{AlzantotSCS2019} and Ilyas18 ~\cite{IlyasEAL18} state-of-the-art black-box attacks. We additionally compare the performance with the $l_{0}$ black-box attack of Su et al. ~\cite{SuVS19}. The rest of the section is organised as follows. \pref{sec:hyperparameters} details the set-up of the conducted experiments and the used hyper-parameter settings of the proposed attacks. \pref{sec:paramstudy} details the conducted parameter study. \pref{sec:peercompare} concludes our attack evaluation by outlining the peer-comparison experiments and their results.
\subsection{Experiments Set-up}
\label{sec:hyperparameters}

\textbf{Hyper-parameters}: The proposed attack makes use of a mutation rate $\mu$ and a radius divider $\beta$ within the mutation and circle objective function respectively. We update the value of the mutation rate $\mu$ throughout the attack process depending on the performance during the current attack. For this, we maintain a variable $pl$ that stores the number of consecutive iterations the objective function does not improve. Once the objective function is improved the value of $pl$ is set to $0$. We employ this update schedule to aid in avoiding local optimums by increasing the exploration of the mutation operator. The final mutation rate $\mu$ is set to $\mu = b \times \frac{pl}{n_{p}}$ where $n_{p}$ an integer. We assign $b$ and $n_{p}$ the values of $0.75$ and $10$ respectively. Evaluating a solution of circle shapes we make use of the parameter $r=\frac{h+w}{\beta}$ which determines the maximum radius of a circle that is placed on the attacked image $x$. In this work, we set the value of $\beta$ to $12$~\cite{ChenLY18,ZouJYZZL19,LiZZL09,BillingsleyLMMG19,LiZLZL09,Li19,LiK14,LiFK11,LiKWTM13,CaoKWL12,CaoKWL14,LiDZZ17,LiKD15,LiKWCR12,LiWKC13,CaoKWLLK15,LiDY18,WuKZLWL15,LiKCLZS12,LiDAY17,LiDZ15,LiXT19,GaoNL19,LiuLC19,LiZ19,KumarBCLB18,CaoWKL11,LiX0WT20,LiuLC20,LiXCT20,WangYLK21,ShanL21,LaiL021,LiLLM21,WuKJLZ17,LiCSY19,LiLDMY20,WuLKZ20,PruvostDLL020}.

\textbf{Experiments Set-Up}: We empirically evaluate the performance of the proposed attack by conducting targeted attacks on 3 state-of-the-art classification models trained on the CIFAR-10 data setting the maximum $l_{\infty}$ distance to $\epsilon=0.05$. To carry out the attack we randomly select 100 correctly-classified images from the CIFAR-10 test-set. For each of the $K-1$ other labels, we aim to find an adversarial example through the optimization of  (\ref{eq:threatmodel}). We evaluate the performance of an attack based on its targeted and untargeted attack success rate (ASR). We denote the targeted ASR as the percentage of adversarial images found that are predicted as the target label. The untargeted ASR represents the percentage of adversarial images found that are not predicted as the true classification label. All attacks are given a computational budget of $10,000$ targeted model queries.

\subsection{Parameter Study}
\label{sec:paramstudy}
\begin{figure*}[!t]
    \centering
    \includegraphics[width=\linewidth]{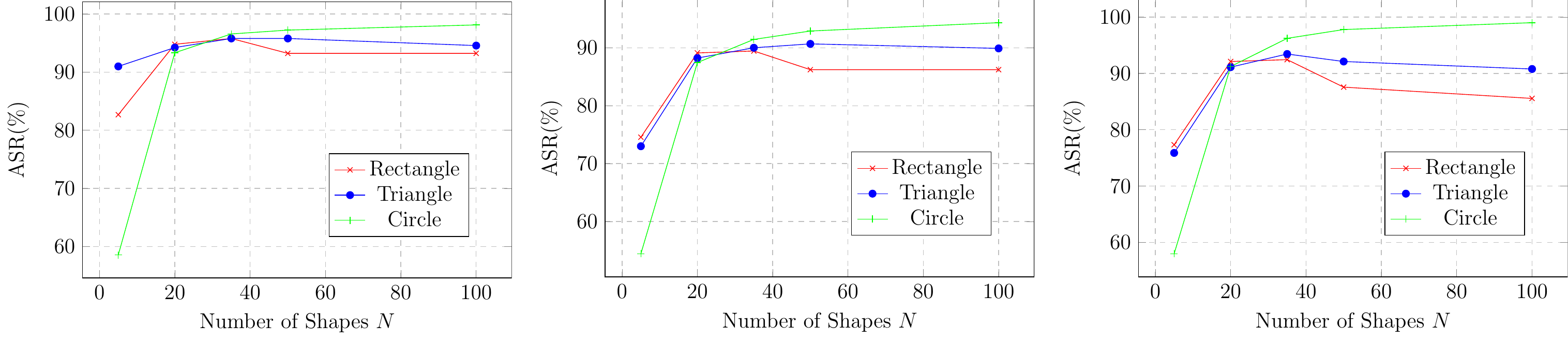}
    \caption{Targeted attack success rate of the proposed attack with varying shape type and number. ~\cite{SpringenbergDBR14} left, ~\cite{LinCY13} middle and ~\cite{SimonyanZ14a} right.}
    \label{fig:cnvasr}
\end{figure*}

\begin{figure*}
    \centering
    \includegraphics[width=\linewidth]{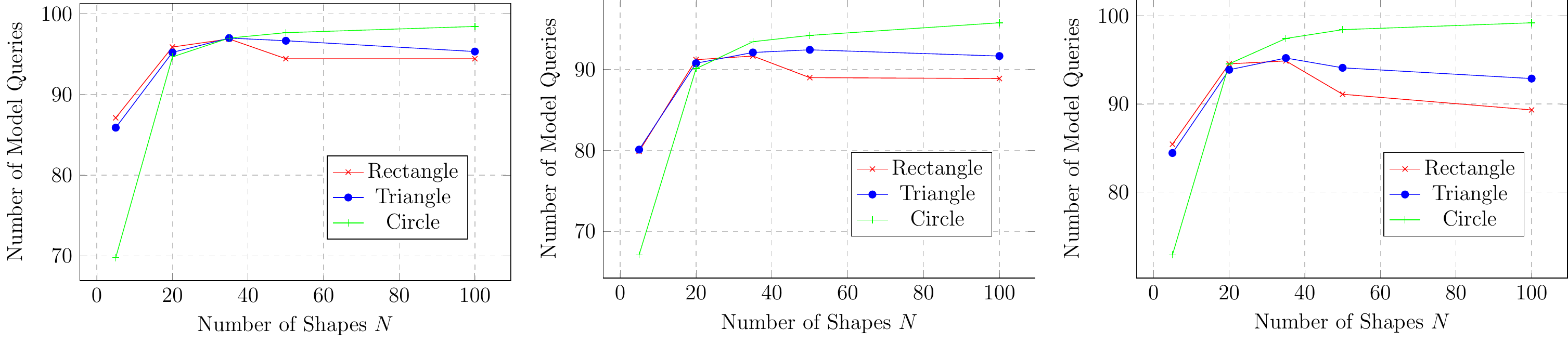}
    \caption{Untargeted attack success rate of the proposed attack with varying shape type and number. ~\cite{SpringenbergDBR14} left, ~\cite{LinCY13} middle and ~\cite{SimonyanZ14a} right.}
    \label{fig:cnvuasr}
\end{figure*}

\begin{figure*}[!t]
    \centering
    \includegraphics[width=\linewidth]{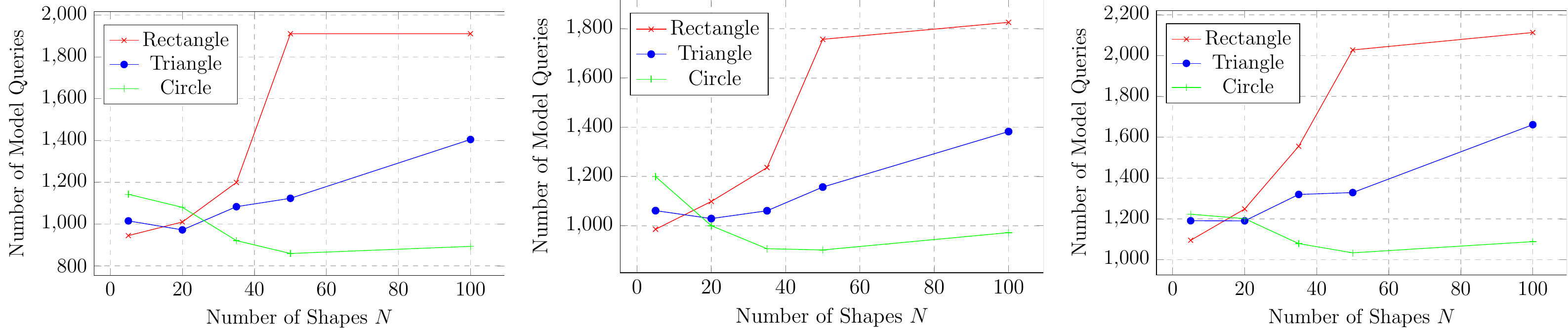}
    \caption{Average number of model queries for the proposed attack to generate successful adversarial examples for the targeted attack. ~\cite{SpringenbergDBR14} left, ~\cite{LinCY13} middle and ~\cite{SimonyanZ14a} right.}
    \label{fig:cnvqueries}
\end{figure*}

We aim to find the optimal type and number of shapes to attack the three used models. To this end, we carry out the targeted attacks outlined in section \ref{sec:hyperparameters} and compare the achieved performance of each parameter setting and give possible reasons for this behaviour.
The plots that are shown in figure \ref{fig:cnvuasr} display the success rate of the attack in locating adversarial examples which are incorrectly classified by the target model. Images classified as the targeted label are included in this data. The results in figure \ref{fig:cnvasr} however only concern adversarial example classified as the targeted label. From the results shown in figures \ref{fig:cnvasr} and \ref{fig:cnvuasr}, we see that varying the type and number of shapes can have a significant impact on the performance of the proposed attack. In particular, we see extreme improvements in performance when increasing the number of circles from 5 to 20, other shapes exhibit similar but less extreme behaviours to increasing the number of shapes. Generally, we see a similar trend of the performance of the proposed attack improving almost exponentially with the increasing number of overlapping shapes. By increasing the number of shapes we enlarge the size of the search space which raises the difficulty of finding an optimum. Figures \ref{fig:cnvasr} and \ref{fig:cnvuasr} display this trade-off as its performance plateaus and also decreases (for triangles and rectangles) when the number of shapes becomes overly large. This prediction is supported in figure \ref{fig:cnvqueries} which shows raising the number of triangles and rectangles used within the proposed attack results in more model queries. This behaviour can be expected as larger search spaces are notoriously more difficult to solve ~\cite{YangTY08} which may require more computational effort. The number of model queries of the proposed attack using rectangles shows the extreme of this. However, we see that increasing the number of circles improves the efficiency of the proposed attack as well as its performance. A possible reason for such behaviour is the impact of adversarial noise constructed from circles has on the targeted model. Shown in algorithm \ref{alg:CircleObjectiveFunction} we make use of the parameter $r$ which determines the maximum radius of a single circle, which is given the value $\frac{h+w}{\beta}$ as outlined in section \ref{sec:hyperparameters}. Applying this limitation on each circle constraints the area that is covered that results in greater sparsity within the generated adversarial noise $\delta$. We do not enforce such a constraint on the size of a triangle or rectangle, as a result, the adversarial noise becomes denser and contains larger areas of overlapping shapes. Exploiting the knowledge of this structure in combination with the results shown in figures \ref{fig:cnvasr} and \ref{fig:cnvqueries} outlines possible weaknesses of the used classification models. More explicitly, from these results, we see that adversarial examples that contain noise within local patches can better exploit the weaknesses of a deep neural network. In future work, we plan to further explore this phenomenon by analysing the impact varying the value of $\beta$ has on the proposed attack performance and efficiency.

\begin{figure}[!t]
    \centering
    \includegraphics[width=.7\linewidth]{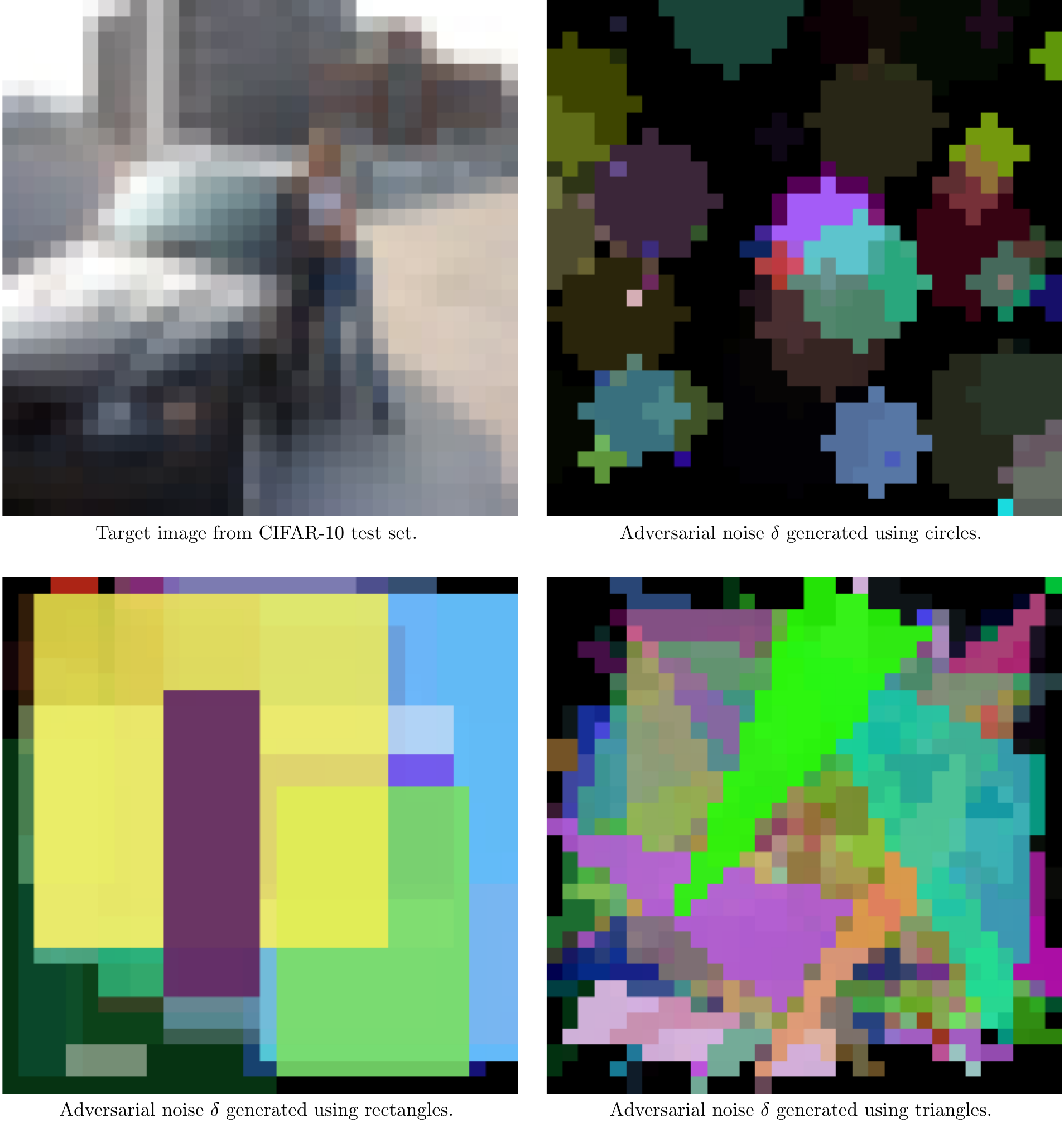}
    \caption{Adversarial Noise $\delta$ generated by attacking ~\cite{SimonyanZ14a} for target image (top left) with $N=50$ before projection.}
    \label{fig:AdvNoiseExample}
\end{figure}

\subsection{Comparison with peer attacks}
\label{sec:peercompare}

\begin{figure*}[!t]
    \centering
    \includegraphics[width=\linewidth]{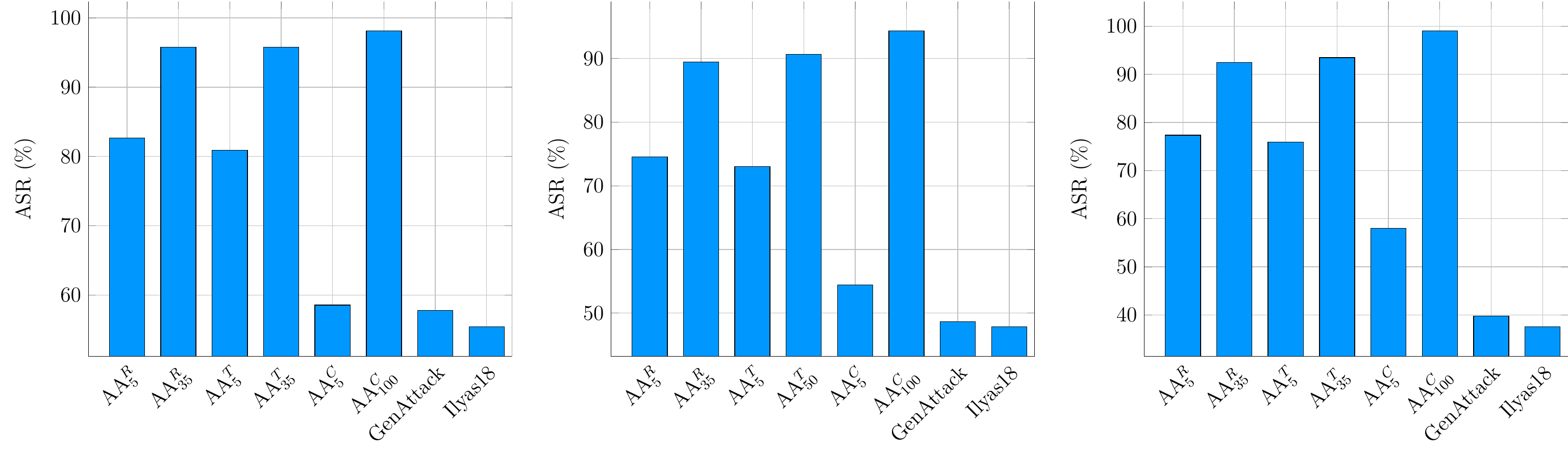}
    \caption{Targeted attack success rate of the proposed, GenAttack and Ilyas18 adversarial attacks. ~\cite{SpringenbergDBR14} left, ~\cite{LinCY13} middle and ~\cite{SimonyanZ14a} right.}
    \label{fig:peerasr}
\end{figure*}

\begin{figure*}[!t]
    \centering
    \includegraphics[width=\linewidth]{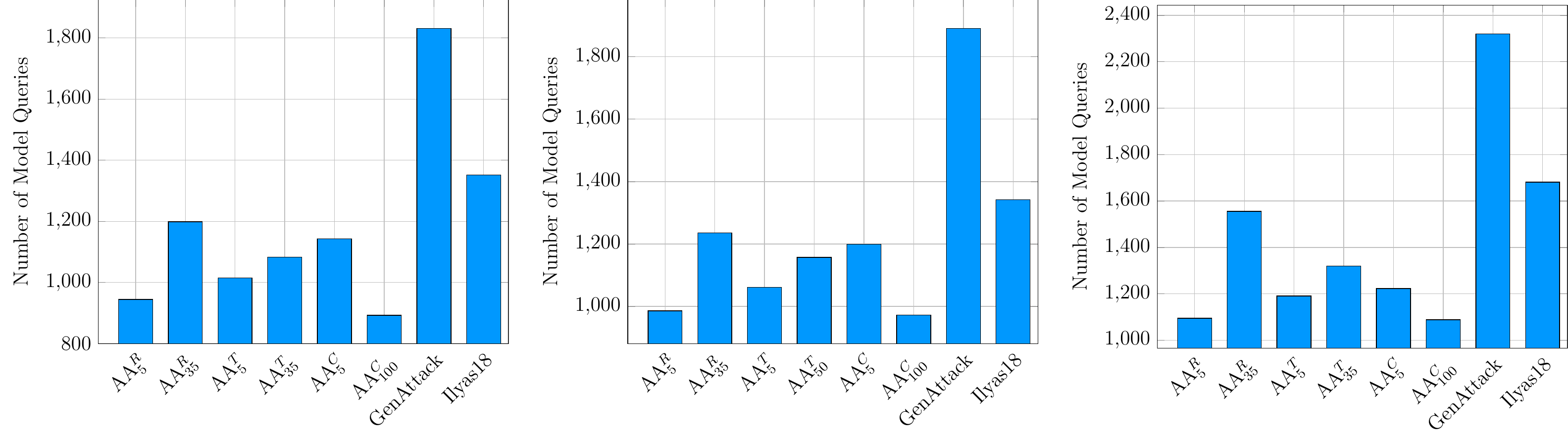}
    \caption{Average number of model queries for the proposed, GenAttack and Ilyas18 adversarial attacks to generate successful adversarial examples.~\cite{SpringenbergDBR14} left, ~\cite{LinCY13} middle and ~\cite{SimonyanZ14a} right.}
    \label{fig:peerquery}
\end{figure*}

To evaluate our proposed attack we first compare its performance to that of GenAttack ~\cite{AlzantotSCS2019}  and Ilyas18 ~\cite{IlyasEAL18} due to their clipping ability to remain within the search bounds $[-\epsilon, \epsilon]$. We further compare with the AutoZoom ~\cite{TuTC0ZYHC19} attack and its performances on the three models. This section concludes with comparisons with the $l_{0}$ One-Pixel-Attack ~\cite{SuVS19}. For full transparency, we compare the performance of the proposed attack with peer algorithms in terms of its best and worse case scenarios. We denote the proposed attack as AA$^{t}_{i}$ where $t$ describes the type (R, T and C) and $i$ the number of shapes.

GenAttack ~\cite{AlzantotSCS2019} is a population-based algorithm that attacks an image by evolving a set of adversarial solutions. To handle the high dimension the original paper uses an embedding space $d^{r}$ and projects into the high dimensional space ($d^{32\times32\times3}$ for CIFAR-10) using bilinear interpolation for solution evaluation. The work of Ilyas et. al ~\cite{IlyasEAL18} also employs this method. For our comparisons we set the embedded dimension $r=8\times8\times3$ as recommended by Tu et. al ~\cite{TuTC0ZYHC19} who reduce the search space $(d^{r}/d)$ by $6.25\% (\frac{192}{3072}\times100)$. Apart from the embedding size, we keep the hyper-parameters of both algorithms constant with that used within their original papers. For both attacks, we employ the (\ref{eq:threatmodel}) as the loss function to be maximized. From the results shown in figures \ref{fig:peerasr} and \ref{fig:peerquery} we that our proposed attack not only outperforms GenAttack and Ilyas18 but also generates adversarial examples using fewer model queries. Despite the varying performance of the proposed attack with changing shape parameters, we see that the proposed attacks worst case still outperforms the two peer attacks. From both sets of results, we see that our proposed attack with the worst parameter settings outperforms both GenAttack and Ilyas18 on all three models. Compared with the proposed attack with its most successful parameter settings, we generate adversarial examples more frequently and with up to $50\%$ fewer queries.

\begin{figure*}[!t]
    \centering
    \includegraphics[width=\linewidth]{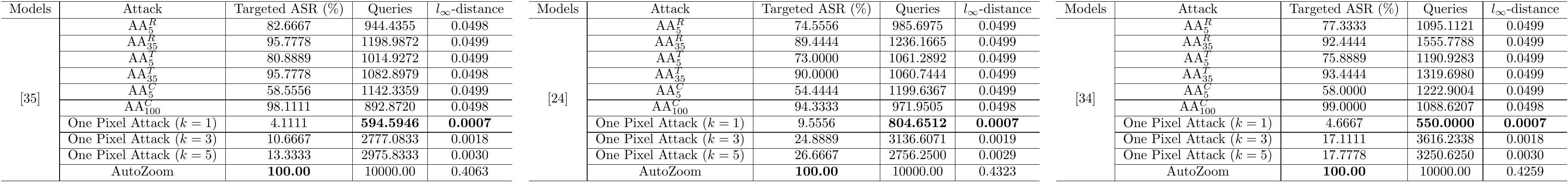}
    \caption{Comparison with AutoZoom ~\cite{TuTC0ZYHC19} and One-Pixel-Attack ~\cite{SuVS19} in terms of average number of model queries, attack success rate, average $l_{\infty}$ distance between the original and adversarial images.}
    \label{fig:autozoom}
\end{figure*}

Whereas both GenAttack ~\cite{AlzantotSCS2019} and Ilyas18 ~\cite{IlyasEAL18} attacks can enforce the $l_{\infty}$ constraint AutoZoom  ~\cite{TuTC0ZYHC19} instead finds an adversarial example $x_{adv}$ by incrementally increasing the search bounds. Once $x_{adv}$ has been found the attack reduces the $\epsilon$ constraint whilst still maintaining a $x_{adv}$ that maximizes the loss function until an adversarial image is generated that is within the original constraints. For our experiments, we employ the constraints used in (\ref{eq:threatmodel}) and terminate the attack once an adversarial example that satisfies the $_{\infty}$ constraints is found. The original AutoZoom paper proposed the use of bilinear interpolation and an Auto-Encoder deep neural network, in our experiments we only make use of bilinear interpolation with $d^{r}=(8\times8\times3)$. We make this decision as both results in the original paper are highly similar which we believe doesn't warrant the use of the computationally expansive Auto-Encoder training. All other algorithm parameters are constant with those used in the original paper. Comparing both attacks in figure \ref{fig:autozoom} we see that the AutoZoom successfully finds an adversarial example for every targeted image. However, even though its high success rate it must be noted that it could not generate a single adversarial example $x_{adv}$ that satisfies the constraint $||x_{adv} - x||_{\infty} \leq 0.05$. This is shown by the attacks average number of model queries (uses the maximum allowed) and the average $l_{\infty}$ distance between the original and adversarial images.

To evaluate the performance of the proposed attack in terms of other attack scenarios we compare with the $l_{0}$ attack proposed by Su et. al ~\cite{SuVS19}, One-Pixel-Attack. To attack a model the algorithm selects $k$ pixels and amends their value by an unconstrained amount. The authors apply a variant of differential evolution ~\cite{StornP97} that only mutates a population of solutions. For our experiments, we keep the attack hyper-parameters constant with those used in the original paper. In comparison with the One-Pixel-Attack shown in figure \ref{fig:autozoom} our proposed attack achieves higher success rates on the three classification models. Where One-Pixel-Attack with $k=1$ generates an adversarial example, it requires a less number of model queries on all three models. This suggests a more efficient attack could first detect sensitive pixels within an image and then exploit them through the addition of some perturbation. These results also demonstrate the difference in difficulty between the two attack scenarios.

\section{Conclusion}
\label{sec:conclusion}

In this work, we proposed a novel black-box attack that uses gradient-free optimization by evolving a series of overlapping transparent shapes which is frequently used within the evolutionary art field. To evaluate our attack we conduct a parameter study that outlines the impact of varying the type and number of used shapes have on its performance. Through this study, we find that a high number of circles are more efficient at attacking classification models within our attack which give hints towards the underlying weaknesses of classification models. In comparison to other attacks in the literature, our proposed attack is able to not only achieve better success rates but also generate adversarial examples using fewer model queries. To the best of our knowledge, this is the first work that evolves a finite set of shapes as an attack method. 

Further developing this work we plan to develop variants of the proposed attack suitable for the partial information and label-only settings outlined in~\cite{IlyasEAL18}. Additionally, to fully evaluate the performance of the proposed attack we plan to conduct studies of the algorithm on a greater range of image classification models. Such models include those that employ state-of-the-art defences and are trained on larger data sets with larger images such as the ImageNet dataset.

\section*{Acknowledgment}
K. Li was supported by UKRI Future Leaders Fellowship (MR/S017062/1), Amazon Research Awards, Royal Society International Exchange Scheme (IES/R2/212077), Alan Turing Fellowship, and EPSRC (2404317).

\bibliographystyle{IEEEtran}
\bibliography{IEEEabrv, references}

\end{document}